%% file: main-manuscript.tex
\documentclass[acmsmall]{acmart}

\usepackage{array}
\usepackage{subcaption}  
\usepackage[most]{tcolorbox}
\usepackage{color}
\usepackage{soul}
\usepackage{multirow}
\usepackage{makecell}
\usepackage{tablefootnote}
\usepackage{multicol}
\usepackage{geometry}
\usepackage{listings}
\usepackage{xcolor}
\usepackage{tabularx}
\usepackage{caption}
\usepackage{boldline}
\usepackage{graphicx}
\usepackage{colortbl}
\usepackage{float}
\usepackage{booktabs}
\usepackage{longtable}

\definecolor{red(ncs)}{rgb}{0.77, 0.01, 0.2}
\definecolor{tabcolor}{rgb}{.126,.126,.126}

\AtBeginDocument{%
  }

\setcopyright{acmlicensed}
\copyrightyear{2018}
\acmYear{2018}
\acmDOI{XXXXXXX.XXXXXXX}
\acmConference[Conference acronym 'XX]{Make sure to enter the correct
  conference title from your rights confirmation email}{June 03--05,
  2018}{Woodstock, NY}
\acmISBN{978-1-4503-XXXX-X/2018/06}

\begin{document}



\title{``I Always Felt that Something Was Wrong.'': Understanding Compliance Risks and Mitigation Strategies when Highly-Skilled Compliance Knowledge Workers Use Large Language Models}

\renewcommand{\shorttitle}{``I Always Felt that Something Was Wrong.''}

\author{Siying HU}
\affiliation{%
  \department{Department of Computer Science}
  \institution{City University of Hong Kong}
  \city{Hong Kong SAR}
  \country{Hong Kong SAR}
}
\email{siyinghu-c@my.cityu.edu.hk}

\author{Piaohong Wang}
\affiliation{%
  \institution{City University of Hong Kong}
  \city{Hong Kong SAR}
  \country{Hong Kong SAR}
}
\email{piaohongwang@gmail.com}

\author{Ka I Chan}
\affiliation{%
  \department{School of Information} 
  \institution{University of Michigan}
  \city{Ann Arbor}
  \country{United States}
}
\email{chankai@umich.edu}

\author{Yaxing Yao}
\affiliation{%
\department{Department of Computer Science}
  \institution{Johns Hopkins University}
  \country{United States}
}
\email{yaxing@jhu.edu}

\author{Zhicong Lu}
\affiliation{%
 \department{Department of Computer Science}
  \institution{George Mason University}
  \city{Fairfax}
  \country{United States}
}
\email{zlu6@gmu.edu}

\renewcommand{\shortauthors}{Hu et al.}

\begin{abstract}
The rapid advancement of Large Language Models (LLMs) has transformed knowledge-intensive has led to its widespread usage by knowledge workers to enhance their productivity. As these professionals handle sensitive information, and the training of text-based GenAI models involves the use of extensive data, there are thus concerns about privacy, security, and broader compliance with regulations and laws. While existing research has addressed privacy and security concerns, the specific compliance risks faced by highly-skilled knowledge workers when using the LLMs, and their mitigation strategies, remain underexplored. As understanding these risks and strategies is crucial for the development of industry-specific compliant LLM mechanisms, this research conducted semi-structured interviews with 24 knowledge workers from knowledge-intensive industries to understand their practices and experiences when integrating LLMs into their workflows. Our research explored how these workers ensure compliance and the resources and challenges they encounter when minimizing risks. Our preliminary findings showed that knowledge workers were concerned about the leakage of sensitive information and took proactive measures such as distorting input data and limiting prompt details to mitigate such risks. Their ability to identify and mitigate risks, however, was significantly hampered by a lack of LLM-specific compliance guidance and training. Our findings highlight the importance of improving knowledge workers' compliance awareness and establishing support systems and compliance cultures within organizations. 

\end{abstract}



\received{20 February 2007}
\received[revised]{12 March 2009}
\received[accepted]{5 June 2009}

\begin{teaserfigure}
  \includegraphics[width=1\textwidth]{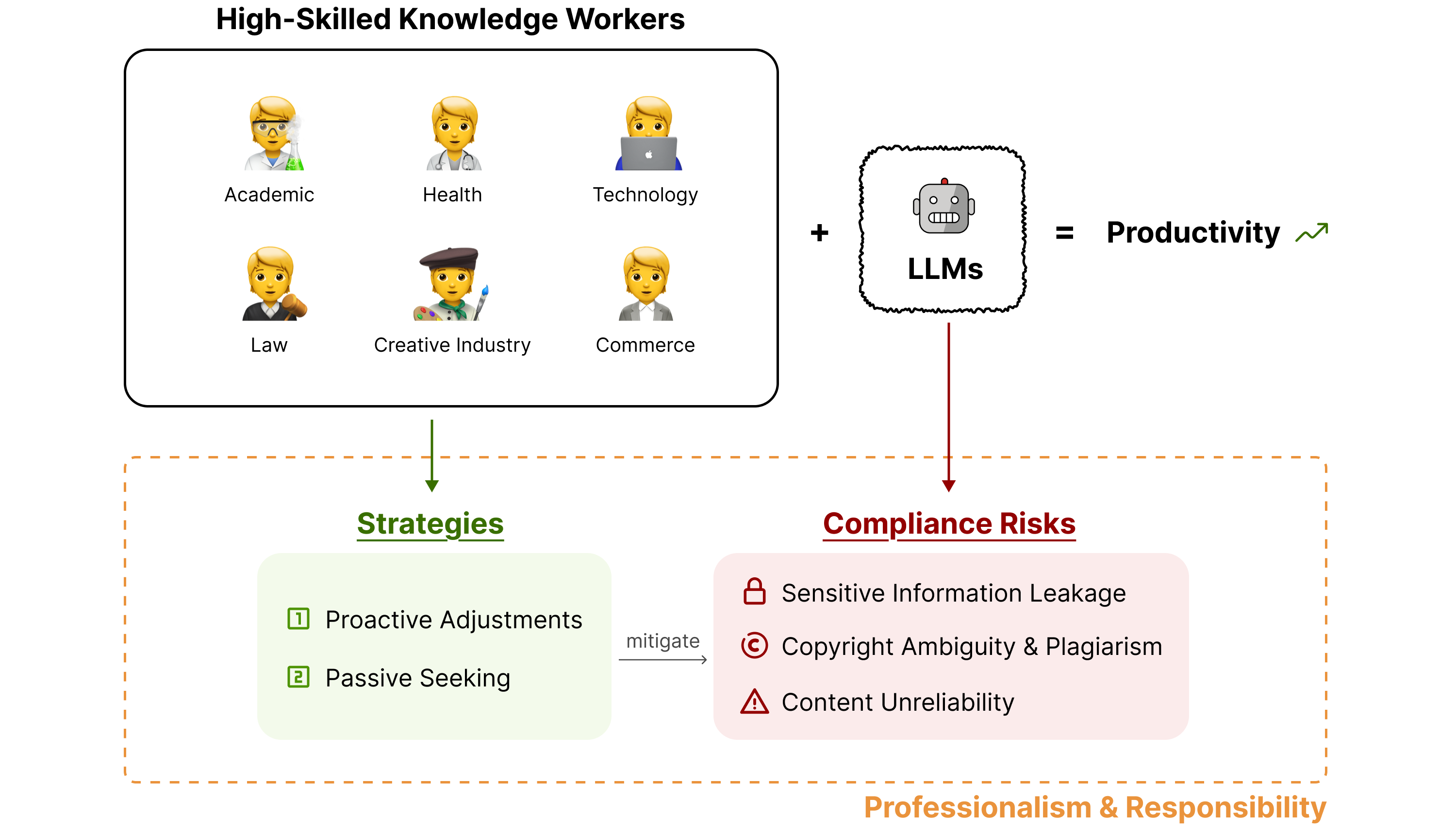}
  \caption{Study Overview. This paper examines the challenges highly-skilled knowledge workers face in ensuring compliance when using Large Language Models (LLMs). We focused on understanding their risk perceptions and mitigation efforts, revealing a critical gap between their ad-hoc strategies and the lack of sufficient organizational and technological support.}
  \Description{This teaser illustrates the core dilemma of our study, capturing the compliance anxieties of highly-skilled knowledge workers who "always felt that something was wrong" when using Large Language Models (LLMs). It depicts their perceived risks (e.g., information security leaks, IP infringement), their ad-hoc mitigation strategies (e.g., distorting input data), and the underlying systemic challenges that undermine their efforts (i.e., model opacity, regulatory lag, and unclear responsibility). Our research unpacks this tension to provide design implications for compliant, human-centered, and responsible AI systems.}
  \label{fig:teaser}
\end{teaserfigure}

\maketitle

\input{sections/01_introduction}
\input{sections/02_related_work}
\input{sections/03_method}
\input{sections/04_findings}
\input{sections/05_discussion}
\input{sections/06_conclusion}




\bibliographystyle{ACM-Reference-Format}
\bibliography{main-references}

\clearpage

\appendix
\section{Examples about Occupation-Specific LLM-driven Applications}
\label{tab:occupation_compliance_example}

\input{tables/ComplianceExamples}
\clearpage

\section{Concerns about Occupational-Specific Compliance of LLM-driven Applications}
\label{tab:occupation_compliance_concerns} 
\input{tables/ComplianceConcerns}

\end{document}

%% file: sections/01_introduction.tex
\section{Introduction}

The rapid development of Large Language Models (LLMs) has transformed knowledge-intensive industries such as research and data analysis, software development and debugging, legal document review, digital marketing, and financial and healthcare decision making \cite{markel2023gpteach,mozannar2024reading,van2024adapted,agarwal2023combining} by enabling such professionals to enhance their individual productivity and elevate the quality and professionalism of their work outcomes \cite{alvesson2001knowledge,markel2023gpteach,constantinides2024implications}. Such professionals, or ``\textit{knowledge workers}'', are individuals who possess specialized knowledge and experience within a particular field and are frequently entrusted with sensitive information \cite{alvesson2001knowledge,drucker1996landmarks}. These knowledge workers leverage their professional expertise to solve practical problems, with the LLMs becoming an increasingly essential tool for them \cite{garneau2021criminelbart,van2024adapted,agarwal2023combining,constantinides2024implications,autor2024applying}. 

While LLMs have offered unprecedented opportunities to enhance worker productivity through rapid information processing, content generation, and data analysis, its reliance on massive amounts of data for training introduces critical challenges related to data security~\cite{wei2023jailbroken}, information accuracy~\cite{kinoshita2022agent}, potential privacy breaches~\cite{zanella2020analyzing,yang2023harnessing}, disinformation~\cite{botha2020fake}, copyright issues~\cite{qu2023new}, and compliance risks with internal company guidelines and GDPR~\cite{baltuttis2024typology,rakova2021responsible,deng2024supporting}. \textit{Compliance risks}, i.e., actions that could have potential for legal, ethical, and reputational repercussions, arise from a variety of sources, including data privacy laws \cite{klymenko2022understanding}, ethical guidelines \cite{trevino1999managing}, and professional standards \cite{fanto2021professionalization} and are particularly pertinent for knowledge workers leveraging new technologies like LLMs \cite{fanto2021professionalization,weidinger2021ethical}. Lawyers may face compliance risks when using LLMs to draft legal documents as the LLMs leaks private information about their clients \cite{cheong2024not}, while doctors using LLMs to diagnose patients may face similar risks if patient data is leaked \cite{qiu2023large}. Recently, Samsung employees were reported to have input confidential information to ChatGPT, including source code for new programs and internal meeting records, while using it to complete work-related tasks \cite{gurman2023samsung}. Given the potential for such compliance breaches, a deeper understanding of the perceived risks and mitigation strategies employed by these knowledge workers is crucial, which can inform the design of responsible and accountable LLM-driven tools for them. 

Although previous studies on LLMs usage have increased our understanding of their technical capabilities and general privacy concerns, several critical questions remain unanswered. From a compliance management perspective, there is limited information on how knowledge workers perceive and navigate compliance risks when using LLMs in their professional contexts, and thus the efficacy of current mitigation strategies remains unclear \cite{liu2024compliance,chandrasekaran2024harnessing}. Additionally, previous studies have not adequately addressed LLMs compliance from the perspective of specialized knowledge work, which relies on complex professional expertise and regulatory understanding to ensure compliance with industry norms and ethical standards \cite{alavi2024knowledge}. Any instance of non-compliance has the potential to result in serious legal, ethical, and damage to professional reputation. A better understanding of these risks and perspectives will help organizations design compliance support systems, develop more effective compliance frameworks, and empower knowledge workers to responsibly integrate LLMs in their workflows while maintaining regulatory compliance in their specialized domains \cite{ulfsnes2024generation}.

Specifically, our research aims to explore the compliance risks that \textit{highly-skilled knowledge workers} perceive when using LLMs in their work and identify the strategies they use to mitigate these risks. We focused on this specific population because these individuals have extensive professional expertise, strong theoretical foundations, extensive practical experience, and exceptional abilities to navigate and resolve complex problems \cite{alvesson2001knowledge,labrague2012knowledge,trevino1999managing}. They possess advanced degrees and demonstrate a high level of cognitive proficiency. These individuals often perform tasks that are subject to strict regulations or industry standards, often featuring a well-defined body of knowledge, educational and certification requirements, and occupational norms \cite{alvesson2001knowledge,labrague2012knowledge,davison2000professional,cheong2024not,shneiderman2020bridging}. Our study specifically focuses on textual work, a core area where LLM-driven tools are currently widely adopted by highly-skilled knowledge workers with significant compliance risks. While LLMs enhances productivity in textual tasks such as drafting, summarizing, and analyzing, its reliance on large amounts of textual data for training raises critical challenges related to data security, privacy breaches, and compliance risks when knowledge workers use LLMs to process text-based information. Highly-skilled knowledge workers frequently rely on LLM-driven tools to process sensitive textual data, making it essential to understand how they perceive and mitigate compliance risks associated with the use of LLMs in their professional work. Within the context of the perspectives of these workers, we sought to answer the following research questions:
 
\begin{itemize}
    \item \textbf{RQ1}: What compliance risks do knowledge workers perceive when using LLMs to complete textual work? 
    
    \item \textbf{RQ2}: How do knowledge workers mitigate compliance risks when using LLMs to complete textual work?

    \item \textbf{RQ3}: What challenges and requirements do knowledge workers have when mitigating compliance risks?

\end{itemize}

To answer these questions, we conducted semi-structured interviews with 24 highly-skilled knowledge workers from knowledge-intensive industries, including law, healthcare, information technology (IT), and academia. Our research concentrated on how these knowledge workers utilized LLMs in their workflows and their attitudes, considerations, and experiences with privacy, security, and sensitive data. We also explored how they ensured compliance with industry norms, company regulations, and professional ethics when using LLMs for work tasks. We found that not all knowledge workers were adequately trained in the compliance risks associated with using LLMs, despite some having previous compliance training. Many intentionally misrepresented their work or limited their use of LLMs as primary mitigation measures. We also found that they needed up-to-date knowledge about model-as-service compliance and the absence of LLMs-specific guidelines and professional codes left compliance open to interpretation and inconsistencies. These factors posed significant challenges for knowledge workers and highlight the need for the identification of best practices, strategies, and systems implication to mitigate compliance risks in professional contexts. 

In summary, our contributions to HCI and CSCW are threefold. 
First, we provide a detailed empirical account of how highly-skilled knowledge workers perceive and navigate compliance risks when integrating LLMs into their professional workflows. We identify the specific challenges they face—stemming from the technology's opacity and regulatory lag—and the ad-hoc mitigation strategies they employ, offering a bottom-up perspective that is currently missing in the literature. 
Second, we highlight the tensions between technological innovation and regulatory lag, emphasizing the need for interdisciplinary approaches to bridging this gap. Finally, we provide design implications for creating systems and facilitating practices that support the ethical, responsible, and compliant use of LLMs, fostering a more sustainable integration of these tools into professional environments. 

%% file: sections/02_related_work.tex
\section{Related Work}
This section builds a three-stage argument that directly addresses how our work differs from and extends prior literature. We first review foundational compliance theories to argue that their traditional focus on user intention is insufficient for the challenges posed by LLMs, where the technology itself is a primary source of risk. We then analyze existing literature on the usage of generative AI tools to pinpoint a critical, unaddressed gap concerning the qualitatively different risks faced by highly skilled knowledge workers. Finally, we critique traditional Compliance Support Systems (CSS) to demonstrate why our findings are not merely new content for old systems, but instead provide the foundational insights for a necessary paradigm shift towards the new paradigm in compliance supports.

\subsection{Understanding Compliance Concepts and the Nature of Knowledge Work}

Within various professional domains, ``\textit{compliance}'' refers to the adherence to specific regulations, industry standards, organizational policies, operational protocols relevant to highly-skilled knowledge workers respective fields \cite{fanto2021professionalization,trevino1999managing}, a concept well-established in fields like cybersecurity  \cite{baltuttis2024typology} and healthcare \cite{ghabayen2023knowledge}. For instance, in cybersecurity, compliance requires strict adherence to data protection laws and security operation procedures \cite{baltuttis2024typology,gupta2023chatgpt}. In healthcare, there is a need to comply with standard precautions to prevent hospital-acquired infections and safeguard  patients and healthcare professionals \cite{ghabayen2023knowledge}. Similarly, in industrial settings, compliance often involves rigorous adherence to health and safety routines and instructions \cite{torp2009influence}. 

``\textit{Compliance risk}'', defined as the potential for legal or regulatory sanctions against organizations and individuals \cite{bamberger2009technologies}, has been a critical concern across various sectors. Research has extensively explored compliance risks in industries such as finance \cite{birindelli2008compliance}, commerce \cite{trautman2018google}, and marketing \cite{kobielieva2018compliance}. Scholars have made significant efforts to mitigate compliance risks in professional settings through regulatory governance \cite{handbook2008governance}, advanced compliance management models \cite{racz2010process}, and novel risk assessment measures \cite{kim2012compliance}. In China, for example, compliance encompasses a range of legal and regulatory standards that vary significantly across industries. For sectors like finance and commerce, compliance is tightly linked to governmental regulations \cite{adam2018emerging}, including Anti-Corruption Laws \cite{rakha2023navigating,adam2018emerging} and financial reporting standards \cite{chen2010impact}. These regulations are designed to ensure transparency and integrity in business operations \cite{sun2024trustllm}. 

While much research has explored compliance from the perspective of organizational security climates on compliance \cite{chan2005perceptions}, the roles of threat appraisal, facilitating conditions, information quality, social influence \cite{pahnila2007employees}, and response efficacy and self-efficacy on the formation of compliance intention and compliance behavior, has garnered increasing attention \cite{herath2009protection}. Concerns about \textit{``compliance behavior'}, defined as employees following organizational rules, regulations, or legal security requirements \cite{warkentin2009behavioral}, has prompted scholars to identify the factors influencing compliance behavior. The Theory of Planned Behavior (TPB) \cite{ajzen1985intentions,ajzen1980understanding} remains the central theoretical framework for analyzing compliance behavior, positing that behavioral intention is the direct antecedent of behavior \cite{ajzen1985intentions,ajzen1991theory}, which is influenced by an individual's attitude towards the behavior, subjective norms (i.e., social pressure), and perceived behavioral control \cite{bass1999individual}. 

Given the inherent complexity and specialized nature of knowledge work, any lapse in compliance can potentially lead to substantial risks and far-reaching consequences for organizations and stakeholders alike \cite{kim2017effect,andreisova2018can}. Moreover, their engagement with rules and regulations often transcends rote obedience, necessitating a nuanced understanding and application of these guidelines grounded in their deep professional knowledge and informed judgment \cite{alvesson2001knowledge,kim2017effect}. Research has indicated that factors such as management and social support significantly correlate with workers' compliance with health and safety guidelines \cite{torp2009influence}. Furthermore, studies such as Putri et al.'s study on COVID-19 have shown a direct link between the level of knowledge and adherence to protocols \cite{putri2023workers}. It is also important to acknowledge the complex, ambiguous, and socially constructed nature of knowledge that underpins the work of these professionals \cite{alvesson2001knowledge}. As compliance decisions are often made using knowledge systems that demand expert judgment \cite{labrague2012knowledge,ghabayen2023knowledge,kim2012compliance} and are shaped by social interactions, professional norms, and their standing within social networks \cite{alvesson2001knowledge}, the risks knowledge workers face not only relate to the adherence of laws but also maintaining a professional reputation.  

In essence, this established body of work paints a clear picture: compliance for highly-skilled knowledge workers is not about rote rule-following, but about the sophisticated application of \textbf{professional judgment} and \textbf{expert intuition} within complex, ambiguous contexts \cite{alvesson2001knowledge, kim2017effect}. This entire paradigm, however, rests upon the assumption that the professional has the ability to scrutinize their tools and environment to make an informed decision. This foundational assumption is shattered by the integration of LLMs. The inherent opacity and generative nature of these `black-box' tools create a direct conflict with the core of knowledge work: a professional is now asked to take responsibility for the outputs of a system whose internal logic they cannot inspect or understand. This creates a novel and deeply personal challenge. The risk is no longer just a potential organizational sanction, but a direct threat to their professional identity and reputation, which are built on the very judgment the LLM now circumvents. Consequently, the gap in the literature is not merely academic; it is a critical failure to address the emergent professional crisis facing knowledge workers. We lack the theoretical framework to understand how these experts navigate compliance when their most trusted tool—their own judgment—is fundamentally undermined by the technology they are compelled to use.

\subsection{Generative AI Tools in Knowledge Work}
\label{supports_knowledgework}

The integration of LLMs as productivity tools in knowledge work is now widespread, supporting specific-tasks from documents drafting and text generation to complex data analysis and automated workflow management \cite{lim2018design,hu2023wizundry,joshi2023repair,10.1145/3290605.3300512,bragg2018sprout}. In specialized fields, tools like Alpha-GPT 2.0 for financial strategy \cite{wang2023alpha} and Deblinder for developer support \cite{cabrera2021discovering}, alongside code assistants like GitHub Copilot \cite{cui2024productivity}, showcase how deeply these systems are becoming embedded in professional practice.

The urgency of this new landscape has prompted organizations to adopt high-level compliance frameworks and guidelines \cite{kim2017effect,koohang2020building,zhang2024enhancing}. These efforts aim to manage regulatory risks by outlining permissible data inputs \cite{hassani2024rethinking}, establishing protocols for verifying outputs \cite{lin2024ethical}, monitoring for disclosures \cite{mokander2024auditing,ferdaus2024towards}, securing information sharing \cite{ferdaus2024towards,chadwick2024assessments}, and offering educational programs \cite{sarker2024llm}. However, while these top-down frameworks are essential for setting organizational policy, they often fail to address the ground-level reality of the risks knowledge workers face. Although the general risks of information leakage are discussed for end users \cite{evripidou2023exploring,mink2023everybody,chan2005perceptions}, and there is a recognized scarcity of resources for responsible computing in industry \cite{deng2024supporting,constantinides2024implications}, these discussions do not fully capture the fundamental nature of the shift LLMs represent.

In response to these emerging threats, organizations have begun adopting high-level compliance frameworks and guidelines \cite{kim2017effect,koohang2020building,zhang2024enhancing}. These efforts aim to manage regulatory risks by outlining permissible data inputs \cite{hassani2024rethinking}, establishing protocols for verifying outputs \cite{lin2024ethical}, monitoring for disclosures \cite{mokander2024auditing,ferdaus2024towards}, securing information sharing \cite{ferdaus2024towards,chadwick2024assessments}, and offering educational programs \cite{sarker2024llm}. Yet, these top-down frameworks, while essential, cannot be fully effective if they do not account for the 'ground truth' of how risks are perceived and navigated by frontline professionals.

We argue this gap exists because the compliance challenges introduced by LLMs are not merely an extension of past technological risks, but are qualitatively different. This novelty stems from three interconnected properties unique to the current ``model-as-a-service'' ecosystem. First is the technology's \textbf{inherent opacity}; knowledge workers cannot inspect the data flows or logic, creating what we term a ``black-box'' risk. Second is the \textbf{ambiguous division of responsibility}; it is unclear whether accountability for a compliance breach lies with the user, their organization, or the LLM provider. Third is the \textbf{generative uncertainty}, where the originality and factual integrity of the output cannot be guaranteed, posing direct threats to intellectual property and professional standards. This confluence of new risks, emerging from a fundamental shift in productivity tools, creates a critical research gap: we do not understand how knowledge workers perceive and attempt to mitigate this new species of risk in their daily practice.

\subsection{Compliance Support Tools with Compliance Intentions}
\label{compliance_behaviours}

In the context of information security, ``\textit{compliance intention}'' is defined as an individual's proactive willingness to protect information and technology resources from potential security threats \cite{bulgurcu2010information,vance2012motivating}. This intention is significantly impacted by behavioral beliefs—such as the belief that adherence to security policies protects privacy and reduces legal risks \cite{ajzen1991theory}—and social pressure from colleagues and superiors \cite{rivis2003social,finlay2002predicting}, a dynamic also analyzed through game-theoretic models \cite{10.1145/3677404.3677423}. Critically, compliance knowledge, an individual's understanding of relevant regulations \cite{desouza2003facilitating,wang2010information}, plays a crucial mediating role. For instance, an emphasis on compliance from superiors (social pressure) \cite{rivis2003social} can motivate employees to acquire relevant knowledge \cite{desouza2003facilitating}, thereby enhancing their ultimate compliance intention \cite{wang2010information}.

Existing Compliance Support Systems (CSS) \cite{10.4018/JGIM.2020040103,kim2017effect} and related tools are designed entirely around this traditional, knowledge-centric paradigm. Functioning as specialized knowledge management systems, their core purpose is to influence knowledge-related behaviors at an organizational level \cite{durcikova2011research,alavi2001knowledge}. They aim to bolster compliance intention by closing knowledge gaps, providing functions like real-time regulatory updates, self-assessment tools, and query support \cite{alavi2001knowledge}. By integrating dispersed organizational knowledge \cite{davenport1998managing,grant1995knowledge} and embedding practical experience into system functionalities \cite{kim2017effect}, these systems are built on a foundational assumption: that providing the right information is sufficient to foster compliant behavior.

However, this information-centric model fundamentally misdiagnoses the problem knowledge workers face with LLMs. As established in the previous section, the challenge is not a simple knowledge deficit, but a profound trust deficit rooted in the technology's opacity, ambiguous responsibility, and generative uncertainty. The critical question for a professional is no longer just ``What is the rule?'' but ``Can I trust this black-box system with my sensitive data and my professional reputation?''. Simply providing more rules or information in a traditional CSS fails to address this core anxiety. This explains the current lack of design guidance for contextual, LLM-empowered compliance tools. What is needed is not just better content for old systems, but a new design philosophy for state-of-art productivity tools. This is where the principles of Human-Centered Responsible AI (RAI) become essential, shifting the goal from merely informing the user to actively building trustworthy, transparent human-AI interactions, and the new paradigm in compliance support. Therefore, our research—by providing deep, bottom-up insights into the specific trust deficits and mitigation needs of knowledge workers—offers the foundational design insights necessary for developing this next generation of RAI toolkits.

%% file: sections/03_method.tex
\section{Study Methodology}
Our study aimed to understand the perceived compliance risks and mitigation strategies employed by highly-skilled knowledge workers (i.e.,  individuals characterized by specific knowledge and skills vital for their roles within their industry, which requires them to adhere to strict codes of conduct and ethical and moral obligations). 

\subsection{Participants}
To build upon prior research on compliance practices within specific sectors \cite{andreisova2018can,atmadja2019proactive,baltuttis2024typology,jin2021lean,sarkar2023exploring,boenisch2021never,10.1145/3613904.3642872}, this study recruited 24 highly-skilled compliance knowledge workers from 8 pivotal sectors, with a balanced representation of 12 females and 12 males (Table \ref{tab:demographic}). Participants were recruited through word-of-mouth and qualified based on their professional status and a minimum of three years of relevant work experience. Participants possessed diverse academic backgrounds, including chemistry, sociology, nursing, computing, communications, and finance. Prior experience with LLMs was not a prerequisite, as a training session was provided to ensure a uniform understanding of generative AI tools and large language models. All participants did, however, have experience using LLM-driven tools such as such as Kimi, Wenxinyiyan, or the POE platform in their work. Due to regional regulations concerning compensation for knowledge workers, participants did not receive remuneration.

\begin{table}[htb]

\caption{Participants Demographics include gender, knowledge of AI, education, occupation, and industry.}
\label{tab:demographic}
\resizebox{\textwidth}{!}{
\begin{tabular}{lcccccc}
\toprule[1.2pt]
\textbf{ID} & \textbf{Age} & \textbf{Gender} & \textbf{AI Knowledge} & \textbf{Education Level}          & \textbf{Occupation} & \textbf{Industry} \\ 
\midrule[1.2pt]
P1          & 30           & Male            & No knowledge             & Doctor of Chemistry               & Research fellow     & Academic          \\
P2          & 31           & Female          & basic                    & Doctor of Social Science          & Lecturer            & Academic          \\
P3          & 30           & Male            & No knowledge             & Bachelor of Laws                  & Public Servant      & Community Service \\
P4          & 30           & Female          & basic                    & Bachelor of Public Administration & Public Servant      & Community Service \\
P5          & 41           & Male          & No knowledge             & Doctor of Medicine                & Surgeon             & Health            \\
P6          & 45           & Female          & No knowledge             & Bachelor of Nursing               & Nurse               & Health            \\
P7          & 33           & Male            & intermediate             & Master of Information Technology  & Software Developer & Technology        \\
P8          & 29           & Male            & intermediate             & Master of Computer Science        & ML Engineer        & Technology        \\
P9          & 28           & Female          & basic                    & Master of Fine Art                & Livestreamer        & Creative Industry \\
P10         & 29           & Female          & basic                    & Bachelor of Communication         & Online Writer      & Creative Industry \\
P11         & 31           & Male            & basic                    & Master of Art                     & Content Creator     & Creative Industry \\
P12         & 26           & Female          & basic                    & Master of Interaction Design      & UX Designer         & Creative Industry \\
P13         & 27           & Male            & No knowledge             & Bachelor of Journalism             & Journalist         & Media \\
P14         & 30           & Male            & basic                    & Master of Laws                    & Lawyer              & Law               \\
P15         & 30           & Female          & basic                    & Master of Laws                    & Lawyer              & Law               \\
P16         & 37           & Female          & No knowledge             & Bachelor of Laws                   & Court Clerk         & Law               \\
P17         & 46           & Male            & basic                    & Master of Laws                    & Court Clerk         & Law               \\
P18         & 35           & Female          & No knowledge             & Bachelor of Financial Accounting    & Auditor             & Commerce          \\
P19         & 29           & Female          & basic                    & Master of Risk Management         & Financial Analyst   & Commerce          \\
P20         & 28           & Male            & basic                    & Master of Finance                 & Trader              & Commerce          \\
P21         & 30           & Male          & No knowledge             & Bachelor of Financial Accounting    & Accounting Manager  & Commerce 
     \\
P22         & 29           & Female          & basic                    & Master of Auditing         & Tax Advisory   & Commerce          \\
P23         & 25           & Male            & basic                    & Master of Finance              & Actuary              & Commerce          \\
P24         & 26           & Female          & No knowledge             & Bachelor of Information Management    & IT auditor  & Commerce 
\\

 \bottomrule[1.2pt]
 \end{tabular}}
\end{table}

This population of participants was selected because they possessed advanced levels of theoretical knowledge and practical skills acquired through formal education, professional training, and  certification or licensing by recognized professional bodies \cite{alvesson2001knowledge}. They interpreted, analyzed, and applied regulations to diverse and often ambiguous situations, requiring a deep understanding of the underlying principles and potential ramifications of non-compliance and carry an explicit responsibility for ensuring compliance. 
They largely worked in sectors where the consequences of non-compliance could be significant, such as healthcare (e.g., physicians or nurses adhering to medical protocols \cite{labrague2012knowledge}), finance (e.g., compliance officers navigating financial regulations), law (e.g., lawyers adhering to legal statutes and ethical codes), engineering (e.g., engineers complying with safety standards), or cybersecurity (e.g., cybersecurity professionals following security protocols and legal frameworks \cite{baltuttis2024typology}). 

Our research employed an ``extreme case sampling'' strategy \cite{suri2011purposeful}. This strategy was used because we hypothesized that the risks and issues arising in domains with the highest compliance pressures (e.g., law, health care, finance), may be similar or more pronounced than those in less regulated industries \cite{alvesson2001knowledge}. For example, those in law and healthcare routinely handle highly sensitive client/patient information, which is a primary area of compliance risk when using LLMs. Such sectors also have different facets of compliance, which would enable us to explore a range of compliance considerations relevant to LLMs usage (e.g., the legal profession is governed by legal statutes and ethical codes, the healthcare industry by medical protocols and patient privacy regulations, and academia by research ethics, data privacy, and intellectual property rights). By examining these ``extreme cases'', we could discover relationships in the data without being restricted by pre-existing assumptions about the specific new compliance risks that LLMs might bring.

\subsection{Procedure}

We conducted semi-structured interviews in Chinese to facilitate clear and nuanced communication, given that the participants were non-native English speakers. The interviews sought to gain insights into participants' perceptions of compliance risks and their mitigation actions while using LLMs for work, inspired by Board's work \cite{board2014professional}. The semi-structured format allowed the researchers to ask predetermined questions to ensure consistency, while also providing the flexibility to ask follow-up questions to delve deeper into participants’ experiences and perspectives. 

Prior to the interviews, participants received an introduction to the fundamental concepts of LLMs, input methods, and expected output effects and content, ensuring a consistent baseline understanding of generative AI tools. We clarified that different industries and organizations may have varying definitions of what constitutes compliance work. Therefore, we explained that \textit{compliance behavior} would refer to actions that aligned with relevant laws, rules, and standards. 

During the interviews, we asked participants about industry-specific practices and experiences they had incorporating LLMs into their workflow and the type of work tasks they performed. We also asked whether there were any specific considerations, attitudes, reasons, or threats related to privacy, security, and sensitive data during this process. We then probed whether they had considered compliance with standard workflows, company regulations, industry norms, and professional ethics. We followed up with questions related to any actions, resources, and challenges they faced when reducing compliance risk in their work and any support they desired when using LLM-driven tools. The interviews were audio recorded for later transcription. The interview process typically lasted between 45 and 60 minutes.

\subsection{Data Collection and Analysis}
After the first author completed the data collection, they transcribed and translated the interview transcripts into English. The interview transcripts and their translations were then cross-checked and confirmed by the second and third authors. Note that all three authors possess an English proficiency level of at least IELTS 6.5 and have completed higher education courses and academic degrees at universities where English is the main language of instruction. 
All data was processed using institution-approved software, was manually correction to ensure accuracy, and then stored on a password-protected, institution-owned computer. All personal information was removed.

Inspired by Merriam and Grenier ~\cite{merriam2019qualitative}, the data was analyzed using a two step open coding process. During the first step, the transcripts were coded line-by-line. Examples of codes included ``delete real name'' and ``damage personal reputation''. This first step was iterative and discussed between two coders to ensure consistency. Two researchers coded 20\% of the data during the first step to generate an initial codebook and then independently coded the rest of the data. They regularly discussed their codes to ensure consistency and added codes to the codebook as needed. Upon completion, they organized the codes into themes. During the second step, the coders synthesized the codes obtained from the first step to extract higher-level themes. The codebook is provided in the supplementary materials for reference. Examples of these higher-level themes included ``Intentional Misleading'' (such as ``misrepresentation of truthful information'' is included) in ``Risk Mitigation Strategies'', ``Professionalism Maintenance'' and ``Avoiding Penalties'' (such as ``avoiding punitive consequences'' and ``personal morality'') in ``Motivators for Compliance''.


%% file: sections/04_findings.tex
\section{Findings}
Herein we identify the perceptions of compliance risk that participants discussed when they integrated LLM into their workflow. We also discuss the challenges and risk mitigation strategies affected by these perceptions and the factors that motivate or inhibit their compliance. Lastly, we present the factors that influence compliance awareness and one's ability to perform risk assessments. The summarization of the time use cases, sensitive data, compliance environment requirements, and compliance concerns for each industry involved in the experiment can be found in supplementary materials \autoref{tab:occupation_compliance_example} and \autoref{tab:occupation_compliance_concerns}. 

\begin{figure}[htbp]
    \centering
    \includegraphics[width=1\linewidth]{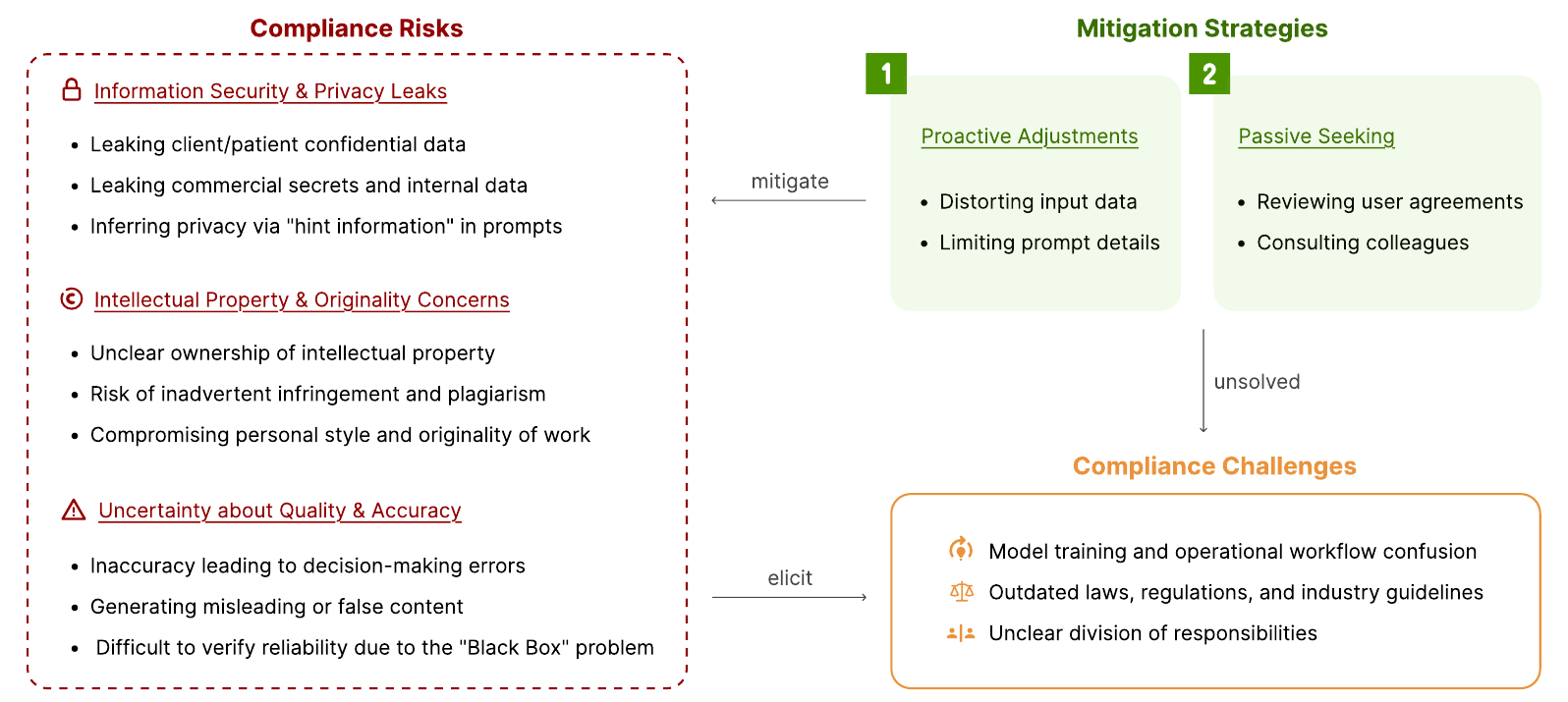}
    \caption{The conceptual framework of compliance risks, mitigation strategies, and challenges derived from our findings. The framework illustrates that knowledge workers perceive three primary \textit{\textbf{Compliance Risks}} (left). In response, they employ two types of \textit{\textbf{Mitigation Strategies}} (top right) to mitigate these risks. Crucially, the effectiveness of these strategies is limited by a set of persistent \textit{\textbf{Compliance Challenges}} (bottom right), highlighting an `unsolved' gap between individual efforts and the need for systemic support.}
    \label{fig:findings}
\end{figure}

\subsection{Perceived Compliance Risks When Using LLM-driven Tools}
Participants perceived several compliance risks when using LLM-driven tools. Here, we define `risks' as the potential, uncertain negative consequences that participants anticipated from using LLMs, such as privacy breaches or intellectual property infringement, which may or may not materialize. These concerns included information security and privacy leaks, challenges to intellectual property, and uncertainty about the quality of generated content.

\subsubsection{Information Security and Privacy Leaks}
A primary concern across all industries was the potential leakage of sensitive information. These concerns, however, manifested themselves differently across different industries. Those in the legal and medical sectors were more concerned about the immediate risk of privacy breaches, i.e., \textit{``leaking customer information is tantamount to professional suicide''} (P15). The work ethic of these individuals drove them to be overly cautious when using new technologies to avoid revealing client or patient information, even in the absence of fully clear regulations. Although the legal profession is governed by legal statutes and ethical codes that necessitate strict confidentiality, there were concerns that confidential customer information entered into LLM-driven tools may be disclosed to third parties or used for model training. Similarly, since the healthcare industry adheres to medical protocols and patient privacy regulations, it must \textit{``ensure that the use of LLM is in accordance with standard operating procedures and ethical practices in the medical industry''}, which often includes strict protections for patient privacy. The use of LLM in these contexts could inadvertently expose sensitive data, leading to severe compliance breaches. Additional examples of industry-specific concerns can be found in \autoref{tab:occupation_compliance_concerns}.



As regulatory environments dictate the laws, regulations, and industry standards that an industry must comply with and the degree of data sensitivity directly affects compliance requirements in terms of data processing and protection, compliance frameworks must be industry-aligned, rather than a one-size-fits-all solution.

Beyond specific industry concerns, participants were also concerned about LLM's handling of both personal and commercial sensitive data. They recognize that LLM's training involves extensive data, raising concerns about privacy and security. Some participants noted the potential for ``hint information '' (P2) within inputs to allow LLM to infer important confidential details that could lead to disclosure violations. Some knowledge workers who were aware of compliance risks also expressed that their awareness of risks was due to their exposure to a large amount of sensitive information in their work. When they were in contact with such sensitive data, they realized the significant responsibility they need to bear, e.g., ``\textit{if these leaks out, it won't be as simple as just accepting the punishment [so] I need to be responsible for this client}'' (P16). As P18 also stated,``\textit{I feel that if I don't follow the rules and regulations they have formulated and something goes wrong when the time comes, who will be responsible?}''. 
Others believed in the value of data, which had led them to recognize the importance of risk sharing, with compliance being a key means to achieve effective risk sharing, i.e., ``\textit{I feel that if I do not follow the rules and regulations they have formulated, something will go wrong when the time comes, and who will be responsible}'' (P18).

\subsubsection{Intellectual Property and Content Originality Concerns}
Another perceived risk related to intellectual property (IP) and the originality of content generated by LLM-driven tools. This was particularly pertinent in creative \cite{ippolito2022creative}, media \cite{hansen2017artificial}, and academic professions. Participants within creative industries and the media felt that there was ambiguity surrounding the role of LLM in the creative process, potentially leading to unclear ownership of intellectual property. They worried that content generated by LLM tool might include copyrighted material that could be improperly attributed or used without a license. Participants expressed uncertainty about their legal obligations in these scenarios. As P10 remarked, \textit{``Maybe I don't ask questions in a good way, so its answers are very monotonous.(...) If I add as more information, then it won't give away more information, I'm not sure if this is legal or not.(...) Like other AI writing fiction tools, I don't dare to use them, I don't know how many of those segments have been used. In the end, I may be spammed by netizens saying that the work is plagiarized. The platform might also deduct my money.''} These comments suggest how metrics and alerts clarifying parameters, such as those concerning intellectual property ownership, copyright compliance, and adherence to academic or professional standards of originality, are needed to ensure that work still conforms to regulations.

In academia, noncompliance with academic standards was perceived to potentially lead to inadvertent plagiarism and intellectual theft. Lecturers, for example, needed to carefully review AI-generated teaching materials to ensure they met academic standards and were not plagiarized. As P2 stated, \textit{``There are also some teachers' teaching courseware that is their own original work, with copyright ownership, and this also needs my additional review and citation. It is also about being responsible for my work and responsible for the students.''}

Beyond IP issues, some participants were concerned that an overreliance on LLM might affect the originality and personal style of their work. While LLM can generate content efficiently, there were questions about whether this content would truly reflect an individual's unique expertise and insights. As P10 said, \textit{``They say it's text generated, but will the generated results be the same as others, and can I use it directly? I don't know about any of this.''}

\subsubsection{Uncertainty about the Quality and Accuracy of Generated Content}
The quality and accuracy of the information generated by LLM was another source of compliance risk because participants were concerned about the potential for errors or inconsistencies when making decisions using such information. Given the ``black box'' nature of generative AI models, it can be difficult to understand the reliability of generated information. Understanding the sources and logic behind the output was difficult and made it difficult to ensure adherence to relevant regulations and professional standards. For professions that require high precision, such as law, healthcare, and auditing, inaccurate information could result in severe consequences. For example, while a lecturer P2 used generative AI to create her teaching slides, she was concerned that compliance issues may arise, e.g., \textit{``It can generate a lot of things, but the content still needs me to carefully review it, because there are many places where the logic is not clear or the content has issues. As a teacher, I also need to ensure the accuracy and originality of my teaching content ... for example, there can't be any bloody violence content displayed in it or content that is not suitable for students to watch.''} (P2).

The potential for LLM to generate misinformation or misleading content was also a significant concern, particularly for journalists, public trustees, and those in public-facing roles. News reports generated by LLM might contain inaccurate information, violating truthfulness and potentially leading to defamation. P17 also emphasized concerns about``misleading suggestions''.

\subsection{Compliance Challenges With LLM-driven Tools}
Beyond perceived risks, LLM introduced several challenges to compliance in knowledge work. In contrast to risks, we define `challenges' as the existing, concrete obstacles and difficulties that currently hinder participants' ability to manage compliance effectively. These related to model training, out-of-date guidelines, and the division of responsibility.

\subsubsection{Model Training and Operational Workflow Confusion}
Participants often lacked a clear understanding of LLM data sources, processing methods, and potential output biases, making it difficult for them to accurately assess potential data leakage and privacy risks. Uncertainty surrounding the training data used by LLM further exacerbated compliance challenges as participants were unaware of the specific sources and content of the datasets used to train LLM models. This lack of transparency raised concerns about potential biases, inaccuracies, or the inclusion of sensitive or copyrighted information. If the training data contained sensitive information, there was also the perception of the potential for a model to inadvertently reproduce or reveal this information in its output. P16 expressed, ``\textit{I can't believe that this thing won't leak to anyone else, so I can only control the source of information and make sure that there will be no problems from my own side}''.

Models-as-service are complex and participants found it difficult to know what, why, and how to conduct safety measures to mitigate compliance risks (N=13). As they lacked a comprehensive understanding of the entire data flow process, they found it difficult to identify and address potential compliance risks. As P13 said, \textit{``the whole process is invisible, and I don't know where the input will go or who will see it''}. Without transparency about how the models arrived at their conclusions, ensuring fairness and compliance became problematic. Participants struggled to understand how the models worked and how results were produced, as well as which security measures should be taken to effectively reduce compliance risks.  For example, P10, a content creator who attempted to use LLM to write some work-related short text segments (snippets), stated that her primary concern was not the specific content or topic of these snippets, but rather the uncertainties surrounding their originality and direct usability when generated by LLM, said, \textit{``they say it's text generated, but will the generated results be the same as others, and can I use it directly? I don't know about any of this.''}

\subsubsection{Out-of-Date Laws, Regulations, and Industry Guidelines}
The rapid development of LLM technology has outpaced the updating of relevant laws, regulations, and industry guidelines, creating substantial compliance uncertainty for participants. This regulatory lag increased the uncertainty faced by participants when ensuring their use of LLM was compliant and would not cause unknown consequences \cite{buiten2019towards}. Some argued that the limitations of LLM services themselves were unclear, making it difficult to determine their scope and restrictions, and they were uncertain how to use them in accordance with workflow or data usage standards. Various laws and regulations did not clearly specify requirements for new technologies, bringing implicit uncertainty to compliance judgments. For example, P18 mentioned that due to the novelty of LLM , there had yet to be a data audit workflow method established within their work handbook, which could pose audit or unknowable risks to auditors and audited parties in the future (e.g., ``\textit{If something is beyond the audited party's knowledge, they won't know, we won't know, nobody in the industry will know, and there won't be any authoritative basis. So maybe we'll only find out in the future that something was wrong or had problems.}''). P18 also said that \textit{``we were only trained on how to use it (LLM ) for some basic tasks, but we don't understand its principles specifically, what can be used, what cannot be used, and how to use them, none of it is explained''}. As P20 stated, \textit{``even though annual updates to exams are just general regulations of the system, they are different from this type of model service. It is unknown if the previous mosaic theory can be applied.''}  The lack of clear guidelines and the rapid evolution of LLM contributed to participants not knowing which aspects of tools to focus on, how to assess their risks, or what controls should be in place. 

We also found that it was difficult for participants to understand how to apply existing rules to new technologies such as language models (N=11). As P19 said, \textit{``for example, we have requirements that you can't use Baidu cloud drive to store data, and you can't upload meeting records to Tencent meetings that are not corporate accounts because of the potential for data leakage. For me, I at least know that I can't use the uploading feature, there's just a clear behavioral red line but the GPT kind of, puts that you don't know how to control the process, you're just conversationally communicating, rather than having that kind of a button to click on.''} At the same time, the lack of clear industry norms and legal guidance made it difficult for participants to which behaviors were compliant and which would be violations.
 
\subsubsection{Unclear Division of Responsibilities When Using LLMs as ``Models-as-a-Service''}
Some participants were unclear about who was responsible for ensuring data protection and compliance. While they often read the Terms of Service for a LLM tool to understand the provider's commitments and liabilities, it was sometimes unclear who would be responsible for breaches. P24 mentioned that the AI tool they usually used was \textit{``recognized within the industry for auditing''} and that it was recognized as having a certain reputation and assurance in terms of data security and compliance. However, P24 was confused about LLM-driven tools' audit processes and methodologies saying \textit{``let's say we have IT general control for system auditing, testing system access, change process, or other control environment and physical security that is somewhat related to this. But if their employees are using ChatGPT, which is not official software, or putting in-house system to complete the work, it is also difficult for us to follow the established audit process to complete the access compliance audit.''}. For this participant, if the LLM service provider failed to protect data, they felt that the responsibility should not fall solely on the user. While service providers often state their data protection responsibilities in their Terms of Service, users may not have a complete understanding of whether the service provider has adequate and effective security measures in place, whether its internal data management processes are compliant, and so on. In the event of a data breach, the responsibility may involve the user's own data security awareness, the service provider's technical protection level, and the agreement between the two parties in the Terms of Service.

Another participant, P20, who was a financial analyst, had easy access to a large amount of commercial transaction data in China listed anti-corruption and anti-commercial bribery as important components of compliance management in Chinese companies that dictated who would be responsible for noncompliance. He was cautious about the use of data involved in his work because of the potential penalties for noncompliance. 
\begin{quote}
    ``\textit{There's also stuff like studying financial law in my major courses. Besides the laws and regulations content we studied when getting our securities qualifications in China, these compliance issues specific to the region will also be covered in additional training at the company, such as the `PRC Anti-Unfair Competition Law' or the `Anti-Corruption Convention'. For example, facing a hefty fine from the China Securities Regulatory Commission (CSRC), the company's management team may face criminal penalties and ultimately cause reputational damage to the company to the point where no one wants to do business with them if there are issues.}''
\end{quote}


\subsection{Risk Mitigation Strategies When Using LLM-driven Tools}
In the face of these perceived risks and challenges, participants highlighted several mitigation strategies they adopted including adapting their workflows, seeking external information to enhance their risk assessments, and maintaining their professionalism to reduce risks.

\subsubsection{Proactively Adjusting Workflows to Reduce Risk}
To mitigate the risk of sensitive information leakage, some knowledge workers used data distortion techniques and limited the details provided in their prompts to LLM-driven tools. Most knowledge workers (N=14) preferred to intentionally misrepresent information to obscure the real information in their work and reduce information disclosure, such as using \textit{pseudonyms} (N=5) instead of real names or intertwining fictional and factual elements (N=4). For example, a researcher (P1) who uses AI tools to reorganize the language of academic writing mentioned, \textit{``I usually delete some of the key content of the original text before using it. I definitely can't let it (LLM) know about keywords, experimental data results, or important arguments, otherwise what if other users find out?''}.  The potential limitation of such intentional misrepresentation is that if the information entered into the LLM tool is overly distorted or key information is deleted, the tool may not be able to accurately understand the user's intent, and the quality and relevance of the results generated may be compromised, reducing the effectiveness of the LLM tool.

Similarly, limiting the amount of detailed information provided in prompts was perceived to reduce the risk of inadvertently revealing confidential details (N=5). These participants were afraid that too many or detailed prompting instructions would reveal their true intentions and easily enable a LLM tool to identify confidential information or easily predict key information based on the context of the inputs. As P2 noted, ``\textit{our industry is very small, and I am worried that there will be too much hint information, one is that the original data will be easily searched by peers, and the other is that not-so-new conclusions will be seen as well.}'' 

\subsubsection{Passively Seeking External Information to Predict and Understand Risks}
Some knowledge workers actively sought out external compliance information from legal resources or colleagues to predict risks and understand compliance responsibilities. Examining relevant legal provisions helped them anticipate potential risk points and take preemptive measures. P16, a law graduate, explained, \textit{``I use various software to check the client agreement to see where our data will be used, how and who can see it, whether they can protect this information and other rules, and it is their responsibility (the service provider) to go through the customer agreement before using the software, and if something goes wrong or the promise is violated ... for example, a WeChat mini-program will provide personal information to third parties to access, so I will consider whether to use it or how to use it'.}'' While seeking external information is a positive risk mitigation measure, reviewing legal texts requires time and expertise, consulting with colleagues or legal professionals can be costly and waste time, and LLM-driven tools' Terms of Service are often lengthy and complex to understand. 

\subsection{Compliance Awareness and Risk Assessments}
The awareness participants had towards compliance issues was influenced by several factors including the (lack of) training they had, their sense of professionalism and feelings of responsibility if issues occurred, and the potential efficiency benefits of using LLM-driven tools.

\subsubsection{The Role of Training on Compliance Risk Assessment}
Compliance perceptions varied across industries due to the training that was available. Some participants were aware of certain compliance risks due to their previous training experiences, including profession-related training, job onboarding, artificial intelligence-related explanations, and other learning channels. This training knowledge made these participants aware of the vulnerabilities that their use of LLM may cause, as well as the foreseeable adverse consequences, which aligns with prior work on compliance with regulations and policies \cite{tahaei2021privacy,leedon}.  

For others, however, there was a clear lack of compliance knowledge training to mitigate compliance risks (N=7). Many of these participants had outdated or insufficient knowledge of new technologies and it was hard for them to keep up with the industry's rapidly iterating knowledge by asking IT staff or reviewing internal guidelines. For example, P15 said \textit{``I learned about this AI from a company training seminar, and besides asking the company's technical staff or compliance department, we had to check the news on the internet, like a lawyer used to generate fake information for a lawsuit or something, to understand the dangers of this model, such as false information, data leakage and so on ... it seems that in addition to knowing this, there is also something that specifically says and privacy and security regulations.''}. They also said, ``\textit{the training we received mainly explains what generative AI is, and how to use it, which is mainly for searching, I think. So far, I understand that there are no uniform industry regulations or processes involved, and most places [companies] should not have this thing, right? Anyway, no one cares about how I use it, and the use of it depends on the individual's choice.''}. 

Others believed that as long as they actively constrained their data-sharing behavior and did not arbitrarily distribute information then it would be difficult to violate compliance regulations. For example, when asked about their practices using LLM, P15 (a lawyer) said ``\textit{we often use voice recorders or mobile phones to record conversations with the parties involved, which facilitates our documentation work. We also use AI tools like iFlytek for transcription and automatic analysis to extract key points and summaries from the dialogue ... GPT is also being used for the same purpose so it is fine.}'' P2 also stated that there were almost no differences between the LLM and AI tools she uses in her daily work and noted how everyone in her industry (education) used them, i.e.,  ``\textit{we often use Grammarly to check the grammar and automatically edit papers. If this application does not comply, is it necessary to retract most papers?}.'' As P11 noted, \textit{``if there are any legal issues, then what is the difference with a cloud disk or photographic software?''}. Participants comfort with existing non-generative tools and their lack of understanding about the differences between AI and LLM-driven tools thus led them to underestimate or misinterpret of the risks associated with LLM tool usage. These comments suggest that these knowledge workers may not be fully aware of the compliance risks associated with using LLM and may not perceive these risks being significant.

\subsubsection{Professionalism and Responsibilities}
Knowledge workers' professional ethics and personal values significantly affected their awareness of compliance, their sense of responsibility for sensitive data, and their perception of the consequences of noncompliance. Some knowledge workers mentioned that using LLM to complete tasks was a morally gray area, lacking corresponding constraint rules and regulatory measures, which could lead one to rely on their personal morality. Unlike the clearly defined ethical standards that knowledge workers in certain fields must follow, the realization of morality when using LLM relied more on knowledge workers' self-discipline and internal self-restraint. 
The absence of specific rules and regulations governing the use of Generative AI in legal practice has prompted legal organizations, such as the one employing Lawyer (P15), to conduct information campaigns aimed at raising awareness among their professionals. In the context of this regulatory ambiguity, Lawyer (P15) highlighted that compliant LLM usage is largely dependent on individual professionalism and ethical judgment. 
As she articulated, 
\begin{quote}
    ``\textit{Which data should be used and which should not, not to the point of being prescribed to this level of detail? Without these specific guidelines, how it is used to get the job done is my freedom. Some people don't really care about this kind of thing when they use it, but there are some rigid people who refuse to use [these tools] altogether just because [they] might leak secrets ... it has to do with their personal character and etiquette.}'' 
\end{quote}

A strong sense of professionalism motivated many participants to consider compliance efforts when using LLM. This included adhering to industry regulations, upholding professional ethics and duties, and ensuring the quality of their work and their reputation. Some expressed that no matter how convenient and powerful LLM services were, they would adhere to professional ethics and never disclose work information in unfamiliar applications. As P23 said, \textit{``it's my habit to check my work equipment and network, and to refuse to use software that is not permitted by the company or specified by the client to process my work content, no matter how awesome that software is.''} 

We also found that participants engaged in compliance behavior to avoid punitive consequences. If there are irregularities in their work, for example, they may face punitive consequences such as fines (N=8), prosecution (N=2), or license revocation (N=9). As P21 mentioned \textit{``Under the terms of the work contract, they can sue us''}. For these participants, their license was their most important asset. As P23 \textit{``If there are any problems that are audited, the most serious revocation of the practice is the revocation of the practice, I'm going to be fired, and then announced, and the whole industry knows about it and I may have to change careers and start all over again.''} Losing a license or facing penalties can not only affect an individual's reputation, but can also lead to serious damage to jobs and livelihoods. Others, howver, mistakenly thought that the ``black box" nature of LLM would absolve them of any legal responsibility. As P9 said, \textit{``I only use snippets so nobody knows if I directly copy it ... only my computer knows''}.

Others engaged in compliance because they were concerned about protecting the reputation of the individual or the organization they worked for. As P19 said, \textit{``customers may not know that their data is being used in GPT and may be held accountable.''} Expert work inherently requires a high level of integrity and if there was a problem with the way LLM assistants were used, it may call into question the reliability and professionalism of an individual or organization. As P14 said \textit{``after all, GPT is not a substitute for real lawyers for legal advice, and if you use the wrong information and evidence they provide, you will be questioned by your clients and ridiculed by your peers''}.

\subsubsection{Efficiency-Compliance Tradeoff}
Within the data, it also became clear that there was a marked tradeoff between leveraging the efficiency benefits of LLM-driven tools and the desire to maintain compliance. Some participants (N=4), for example, discussed how they were driven by the behavior of colleagues to use LLM to improve work efficiency and thus overlooked compliance considerations. As P9 said, ``\textit{I didn't expect to use this, but I saw that my colleagues around me were using it, and more and more, everyone said yes, and it was indeed better than our usual writing, and I didn't pay special attention to what the process or regulations were not in line with, so I used it too}''. Productivity gains are an important benefit but must be weighed against privacy and security risks. LLM can streamline workflows but also introduce vulnerabilities if sensitive data is exposed without safeguards. Some participants noted how speed alone should not compromise critical information. More robust tools are needed to proactively identify privacy and security issues when conducting noncompliance work. 

Excessive precautions, however, can also overburden users and clash with efficient workflows. Strong protections increase complexity, thus slowing processes, while at the same time, basic due diligence prevents noncompliance. P7 said, \textit{``As a tool, it definitely improves my work efficiency. Because I'm really not good at writing. If I need to spend more time to repeatedly filter the unusable information, I will think that the time spent is not worth it, and this will change from reducing the burden to increasing the burden. This tool has no meaning of existence, it defeats the purpose.''}. Navigating this efficiency-compliance tradeoff thus demands guidance on its impact on workflows and priorities.

%% file: sections/05_discussion.tex
\section{Discussion}
This research examines the compliance risks associated with LLM in professional practices. 
For \textbf{RQ1}, we found that knowledge workers primarily perceived the following key compliance risks: the risk of leakage of sensitive personal and commercial information, especially concerns about violating client and patient confidentiality in the legal and medical fields; issues of intellectual property and content originality, 
and uncertainty about the quality and accuracy of LLM-generated content.
For \textbf{RQ2}, we found that knowledge workers primarily adopted the following proactive risk mitigation strategies: actively adjusting their workflows to reduce risks, such as by distorting input data and limiting the detail of prompts to decrease the possibility of sensitive information leakage; many knowledge workers tended to intentionally misrepresent information, using pseudonyms or intertwining fictional and factual elements to obscure real data. 
For \textbf{RQ3}, we found there is a lack of clear compliance guidance and training tailored to LLM and knowledge workers themselves, which makes it difficult for them to accurately assess and address potential risks; a lack of clear understanding of LLMs' data sources, processing methods, and potential biases, hindering the implementation of effective risk mitigation measures; existing laws, regulations, and industry guidelines lagging behind the rapid development of LLM technology, creating uncertainty in compliance judgements; and an unclear division of responsibilities when using LLM as a `models-as-a-service'. 

Our research primarily address the impact of LLMs on professional workflows across various industries and their implications for legal regulations. We synthesize approaches to enhance knowledge workers' awareness, motivation, and capabilities to mitigate compliance risks when integrating LLMs into their workflows. Our focus remains on compliance-specific recommendations, aiming to provide pragmatic guidance for managing the regulatory challenges posed by LLM adoption in professional settings.

To address the multifaceted compliance challenges identified in our findings—namely the lack of guidelines, model opacity, and unclear responsibilities—we propose a set of actionable design recommendations. These recommendations are not arbitrary; they are systematically grounded in the principles of Responsible AI \cite{gadekallu2025framework}. We adopt the RAI framework because the challenges are not merely technical, but deeply socio-technical, involving human behavior, organizational culture, and ethical considerations. The RAI framework provides a holistic lens, emphasizing critical dimensions such as Transparency, Governance, Ethics, Safety, and Accountability. Accordingly, our discussion is structured around four key strategic goals derived from this framework: (1) Enhancing Awareness in section 5.1, (2) Eliciting Motivation in section 5.2, (3) Empowering Ability in section 5.3, and (4) Fostering an Organizational Approach in section 5.4. Each of these goals directly tackles the underlying challenges and paves the way for a more compliant and responsible use of LLMs in the workplace.

\begin{figure}[htbp]
    \centering
    \includegraphics[width=1\linewidth]{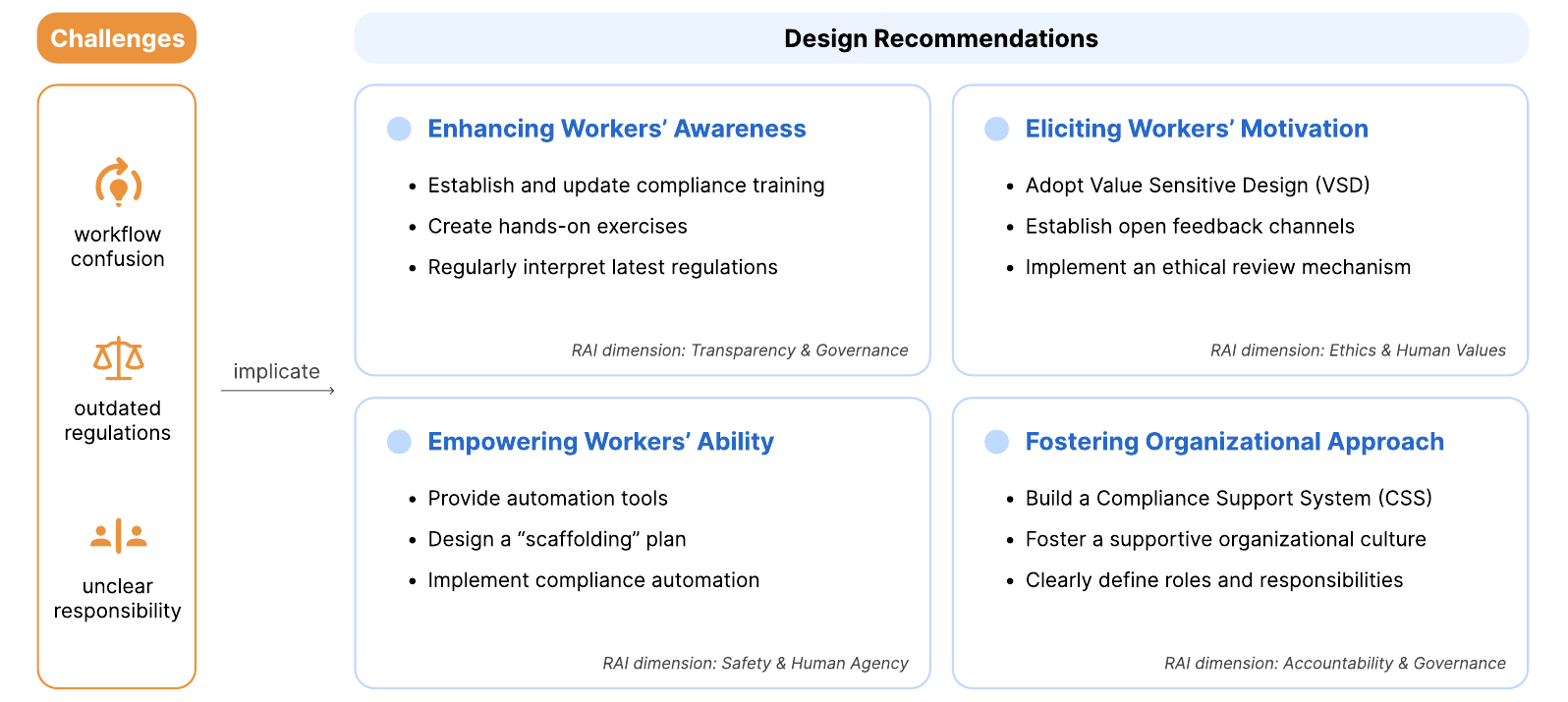}
    \caption{Our design framework, illustrating four sets of recommendations to address the compliance challenges. The framework shows how the \textit{\textbf{Challenges}} (left) identified in our study implicate the need for a multi-faceted approach. The \textit{\textbf{Design Recommendations}} (right) are organized into four key areas: (1) Enhancing Workers' Awareness, (2) Eliciting Workers' Motivation, (3) Empowering Workers' Ability, and (4) Fostering an Organizational Approach. Each area is linked to a specific Responsible AI (RAI) dimension, providing actionable pathways for creating more trustworthy and compliant systems.}
    \label{fig:findings}
\end{figure}

\subsection{Enhancing Knowledge Workers' Awareness of Compliance Risks}
This research revealed that although some knowledge workers possess a certain level of awareness regarding compliance risks, they generally lack sufficient training on the specific compliance risks associated with the use of LLM. This lack of understanding may lead to conscious information bias, such as using pseudonyms, or limiting the use of prompts to mitigate risks. Therefore, up-to-date compliance knowledge concerning model-as-a-service is crucial for knowledge workers, and there is an urgent need for more targeted education and training. To this end, we propose the following potential methods to enhance knowledge workers' awareness of compliance risks. 

Firstly, generative AI models may produce inaccurate or incomplete results, posing potential risks in professional fields such as finance and medical diagnostics. To mitigate such risks, knowledge workers must fully understand the compliance threats that integrating LLM into their workflows may trigger. Organizations should thus establish and regularly update practical and effective compliance training programs, tailoring the training content to the specific needs of different positions and industries \cite{wagman2025generative}. Such training should use industry-related real or simulated case studies to identify compliance risk points in the LLM usage process \cite{deng2024supporting}. For example, in the healthcare sector, practitioners should be reminded to consider the type of hospital from which data is obtained (public or private) and the differences between public and private health insurance companies. 
To more effectively enhance knowledge workers’ awareness of compliance risks when using LLM, we suggest creating practical exercises simulating LLM usage scenarios, embedding potential compliance risks within them, and conducting post-exercise reflection and discussion.

Second, organizations should regularly interpret the latest laws and regulations related to AI, data privacy, and intellectual property, and invite experts to provide detailed explanations of the implications of these laws and regulations \cite{kyi2025governance}. For example, the EU's Artificial Intelligence Act identifies employment and the labor market as high-risk areas \cite{jarota2023artificial}. Training should elaborate on the potential privacy and security risks in the LLM data processing, emphasize security issues in prompt engineering, and introduce best practices for data protection. 

Third, best practices in organizational ethics education and training should be adopted, such as emphasizing the interactivity and participation of training, integrating compliance training with organizational values, using diverse training formats, establishing training effectiveness evaluation mechanisms, and conducting continuous compliance communication and reminders \cite{sekerka2009organizational,sadek2024guidelines,deng2024supporting,constantinides2024rai}. Companies should establish practical and effective compliance training programs that are updated in line with the latest regulations and laws. 

Through these specific measures, the compliance risk awareness of knowledge workers when using LLM can be more effectively enhanced, and they can be guided to take appropriate risk mitigation measures.

\subsection{Eliciting Knowledge Workers' Motivation for Proactive Compliance Mitigation}
\label{tab:work_motivation}

The results of this study show that knowledge workers' professional ethics and personal values significantly affect their awareness of compliance, their sense of responsibility for sensitive data, and their perception of the consequences of non-compliance, all of which affect their compliance behavior at the subconscious level. 

As Meyerson's concept of ``tempered radicals'' elucidates, employees can foster organizational social responsibility through subtle, sustained efforts \cite{meyerson2004tempered}. For knowledge workers, their emphasis on professional ethics and compliance can be channeled into a form of ``tempered radicalism,'' enabling them to sensitively identify potential ethical and compliance risks in LLM usage. They can then construct their work in a manner that both upholds their professional integrity and garners organizational attention, such as by highlighting how compliance enhances service quality and safeguards organizational reputation, thereby elevating compliance issues to a priority within the organization. Moreover, Rakova et al.'s research investigated the factors driving organizational practice transformation \cite{rakova2021responsible}. Their findings underscored the critical role of frontline practitioners' perspectives and initiative in the implementation of responsible AI. Drawing on this insight, we can encourage knowledge workers to integrate their professional ethics into daily operations, proactively engage in refining compliance processes, and actively share compliance challenges and best practices encountered during LLM utilization. In this way, individual moral impetus can gradually influence organizational culture, fostering a stronger compliance awareness and more effective risk management mechanisms.

As we found in our study, to foster proactive risk mitigation, many knowledge workers required clear data flow and parameters of AI tools; therefore, design transparency and accountability should be adhered to, and intrinsic motivation should be cultivated through ethical values \cite{atmadja2019proactive}. Building on pro-social design considerations that incentivize practitioners to address AI-exacerbated threats \cite{lee2024don}, we propose a potential approach that encourages pro-social behavior by viewing compliance as a positive pro-social contribution. By establishing open feedback channels to facilitate effective communication between knowledge workers and technical and legal teams about compliance issues, knowledge workers can foster a shared sense of accountability. To ensure that compliance approaches adequately consider the personal needs, values, and trade-offs of knowledge workers, it is recommended to implement ethical review mechanisms that encourage participation from other knowledge workers within the same industry or community. These reviews offer valuable opportunities for knowledge workers to learn from peers through practice communities, driving improvements in their safety and privacy practices on an intrinsic level. 

The integration of Value Sensitive Design \cite{sadek2024guidelines} and Responsible AI \cite{rakova2021responsible,constantinides2024rai} emphasizes incorporating core human values into the technology design process. From the perspective of individual motivation, when knowledge workers realize that their use of LLM aligns with ethical and social responsibilities, they are more likely to develop an intrinsic sense of identification and willingness to comply. If the regulatory boundaries are unclear and responsibilities are not well-defined, knowledge workers may find it difficult to recognize the specific compliance risks associated with using LLM, and may also lack the motivation to take proactive mitigating measures \cite{10.1145/3630106.3658998}. By emphasizing the alignment between principles such as fairness, privacy, informed consent, and professional ethics and values, knowledge workers can effectively inspire their intrinsic compliance motivation. Establishing positive feedback mechanisms that recognize and affirm the compliant behaviors of knowledge workers also helps reinforce their intrinsic sense of accomplishment and motivation for continued compliance.

\subsection{Empowering Knowledge Worker's Ability to Use Risk Mitigation Strategies}

Knowledge workers require more resources to use risk mitigation measures. The absence of guidance hampers their ability to assess risk hazards and to comply with model-as-service requirements. Detailed role-based procedures and automation tools are needed to empower effective risk assessment and mitigation \cite{zakariarisk,kim2016integrating}. Previous studies have demonstrated that a well-designed scaffolding plan can establish a systematic workflow that adheres to regulations and standards \cite{zakariarisk}. The scaffolding can be customized according to industry-specific model documentation templates and audit programs, effectively structuring processes for professional compliance work protocols \cite{kim2016integrating}. It can also provide role-specific standard operating procedures, control matrices, and compliance checklists as reference material to ensure a consistent and compliant approach in the workplace.

Implementing compliance automation and tooling can also reduce the manual effort required for repetitive compliance tasks such as documentation, testing, and monitoring model drift over time \cite{bedi2020basic}. This will ensure consistency and accuracy in applying controls while minimizing human errors \cite{dzuranin2016current}. 
Companies can leverage model monitoring tools to continuously track performance metrics and detect anomalies in LLMs, and utilize testing frameworks that automatically execute profession-specific checklists on different versions of these models. They can also streamline the process by integrating documentation generation tools that extract and update metadata at each stage.

LLM models can also be fine-tuned to leverage the abundant professional knowledge that knowledge workers have. Techniques like few-shot learning could be used to ensure good performance using only limited input data. 
Considering the costs of fine-tuning LLM, a feasible strategy is to anonymize or de-sensitize user input data before providing it to LLM, which could also greatly alleviate professional users' anxiety about data security.

\subsection{Fostering an Organizational Approach Towards Compliance Risks}
While it is crucial to inspire the proactivity of knowledge workers, more systematic and effective compliance risk management also requires organizations to build corresponding support systems and a cultural atmosphere at the organizational level. Drawing from Shneiderman et.al.,'s guidelines for constructing reliable, secure, and trustworthy human-centered AI systems \cite{shneiderman2020bridging}, organizations should strive to bridge the gap between ethical principles and practical implementation. The construction of a compliance culture is a complex and incremental process that requires organizational leaders to explicitly advocate for and embody these principles, while creating an environment that supports employee engagement and contribution. Merely issuing compliance policies is insufficient; organizations must actively identify and eliminate barriers to cultural transformation, such as potential retaliation for reporting and indifference to ethical discussions. In the absence of effective communication and transparent feedback mechanisms, encouraging the reporting of violations may become an empty gesture, potentially leading to negative speculation and distrust among employees.

Building and effectively leveraging a compliance support system (CSS) is a critical part of developing an organization's approach to compliance risk.Future CSSs should automatically generate compliance guidelines and operational processes for LLM use based on the latest laws and regulations, such as the EU AI Act, and industry standards. Frameworks and tools, such as StakeLinker and FillGen, demonstrate how AI technologies, specifically LLM, can be leveraged to streamline the impact assessment process, including risk identification and the generation of mitigation measures \cite{bogucka2024ai}. Deng et al.'s impact assessment methods and tools could also be applied to assess the compliance risks of LLM in specific scenarios, helping knowledge workers to more systematically identify potential legal and ethical issues \cite{deng2024supporting}. Bogucka et al.'s research that proposed a responsible AI framework for pre-populating impact assessment reports could also be leveraged \cite{10.1109/MIC.2024.3451351}. Moving forward, it will thus be necessary to combine traditional compliance management concepts with emerging technological capabilities to empower knowledge workers, improve their risk assessment and mitigation capabilities, and ultimately build an organization-wide compliance risk defense line.

\subsection{Ambiguity of Knowledge Work with Compliance Risks}
This research focuses on the unique compliance risks faced by highly-skilled knowledge workers when using LLM. As highly-skilled knowledge workers need to apply their professional knowledge and judgment to interpret and apply intricate regulations, their work often involves ethical responsibilities, especially when handling sensitive information. These inherent complexities, combined with the uncertainties introduced by LLM technology, create unique compliance challenges. For example, the operational principles and data processing flows of LLM are often opaque to users, making it difficult for knowledge workers to assess potential data leakage and privacy risks. Furthermore, due to the rapid development of LLMs, relevant laws, regulations, and industry guidelines often lag behind, further increasing compliance uncertainty. 

As previously research stated, the complexity of knowledge, the ambiguity of knowledge, and the socially constructed nature of knowledge further exacerbate compliance risks in knowledge work \cite{alvesson2001knowledge,wagman2025generative}. Highly-skilled compliance is not just about memorizing rules but also about understanding the principles behind the rules and applying professional knowledge and skills to practice compliance. As we found our research, when facing complex regulations and ambiguous situations, the use of LLMs may blur the boundaries of applying professional knowledge, making it difficult for knowledge workers to judge the reliability and compliance of LLM-generated content. Due to the black-box nature of LLM, knowledge workers may find it challenging to understand the sources and logic of their output, thereby making it difficult to ensure their adherence to relevant regulations and professional standards. Moreover, the cognition and practice of compliance are also influenced by social interactions and professional norms, while the relevant social norms and best practices for LLM as an emerging tool are still in the process of being formed.

The rapid advancement of LLM technology has significantly outpaced the updating of existing laws and regulations, leading to substantial compliance uncertainty for highly-skilled knowledge workers. This uncertainty stems not only from regulatory lag but also from the ambiguity surrounding the capabilities and limitations of LLMs  themselves. Knowledge workers often lack a clear understanding of LLM data sources, processing methods, and potential output biases, making it difficult to accurately assess the compliance risks associated with their use in work. As highlighted in our findings, many knowledge workers pointed out that current laws, regulations, and professional ethical standards do not clearly define specific usage standards and compliance requirements for emerging LLM collaborative technologies.

\subsection{Transferability of Our Findings}
Our findings suggested that even in highly specialized fields, knowledge workers commonly face insufficient awareness of the boundaries and potential biases of LLM , which directly impacts their assessment of compliance risks. Furthermore, we observed that when confronted with compliance uncertainties arising from emerging technologies, professionals tend to rely on existing professional ethics and industry norms for self-regulation, although a clear consensus on the attribution of responsibility and verification standards for LLM-generated content has yet to be formed. While this study's sample size and industry distribution have limitations, its findings may offer some reference value for understanding other fields or groups of knowledge workers with similar characteristics. For instance, industries that also require handling large amounts of sensitive information, face complex and evolving regulations, and heavily rely on professional expertise and judgment may also encounter similar challenges in identifying compliance risks, applying rules, and determining responsibilities when using LLMs , which also found in previously work \cite{bulgurcu2010information,agarwal2023combining}. 

It is crucial to emphasize that the core focus of this research is on ``highly-skilled knowledge workers.'' We chose them as our research subjects because they bear the critical responsibility of ensuring organizational operations comply with laws, regulations, industry standards, and internal policies within their respective professional domains. Their work demands not only profound professional knowledge but also a high degree of judgment and ethical awareness to address complex compliance issues. Notably, compliance work is often directly linked to an organization's reputation, legal liabilities, and customer trust. Exploring the impact of LLMs in responsible AI and trustworthy holds significant practical relevance \cite{rakova2021responsible,constantinides2024implications,constantinides2024rai}. 

To further enhance the generalizability of our research findings, future studies could consider expanding the industry coverage of participants, for example, by including other knowledge-intensive industries such as biotechnology or engineering, and by conducting cross-industry comparative analyses. Additionally, adopting more diverse research methods, such as incorporating case studies, quantitative analyses, and examining knowledge workers in organizations of different sizes and cultural backgrounds, could contribute to a more comprehensive understanding of the complex relationship between LLM use and compliance risks, ultimately providing a theoretical basis for developing more universally applicable compliance guidance and practices \cite{elsayed2023responsible}.

\subsection{Limitations and Future Work}
t is crucial to understand that this study's aim was exploratory, focusing on the initial perceptions and strategies of a specific group of highly-skilled compliance knowledge workers in selected industries. It did not intend to provide a generalized assessment of compliance risks across all professions or sectors. This qualitative approach was chosen to deeply explore the nuanced perceptions of compliance risks and the subjective mitigation strategies adopted by these professionals in their real-world work scenarios. This methodology allowed for the collection of rich, exploratory data regarding their understanding and adaptation to emerging technologies like LLMs  within the context of compliance. As such, we acknowledge that the experiences of professionals in the sectors we interviewed may not be directly generalizable to all other professions employing LLM or sectors with different regulatory landscapes and operational characteristics. Future research should address this limitation by expanding the participant pool to include professionals from a wider array of knowledge-intensive industries, such as data science, engineering, and information technology, to provide a more comprehensive understanding of compliance risks associated with LLM adoption. This would allow for a comparative analysis and sector-specific best practices for the use of LLMs.

We also relied on participants' subjectivity perceptions, which may not accurately reflect actual risks. Consequently, we cannot draw definitive conclusions about real compliance risks. We can only report on knowledge workers' attitudes and awareness. Future research should address this by combining a technical analysis of LLM usage with comprehensive legal and regulatory assessments, and objective knowledge tests for knowledge workers. This multi-faceted approach would allow for a more holistic evaluation of the gap between perceived and actual risks. Additionally, longitudinal studies could track how these risks evolve with advancing LLM technologies.

The in-depth exploration of AI agents and agentic collaboration within knowledge worker workflows holds paramount significance, as it promises to not only enhance work efficiency and quality but also reshape the organization and execution of knowledge work. Future research should focus on investigating the novel compliance risks and corresponding mitigation strategies that knowledge workers may encounter when interfacing with AI agents and multi-agent collaborative work scenarios. The ultimate goal will be to ensure that knowledge workers can fully leverage the efficiency gains and enhanced work quality brought by AI agents and agentic collaboration within a secure and compliant environment.

\vspace{-0.5em}

%% file: sections/06_conclusion.tex
\section{Conclusion}
This research explored the compliance risks highly-skilled knowledge workers perceive when using LLM and the strategies they adopt to mitigate them. Through semi-structured interviews with highly-skilled knowledge workers, we found that while LLM enhances work efficiency, their key concerns about compliance risks when using LLM include potential privacy breaches and regulatory violations. To address these risks, workers have taken proactive mitigation measures such as intentionally misrepresenting input data, limiting the detail of prompts, and seeking external compliance guidance. However, for many interviewees, the lack of industry-specific compliance guidance and training tailored to LLM and the ambiguous nature of domain knowledge make it challenging for knowledge workers to fully understand and effectively address emerging compliance risks. Simultaneously, the rapid development of LLM technologies and the lag in relevant regulations have further increased uncertainties in compliance work. Our findings underscore the importance of enhancing knowledge workers' compliance awareness, motivation, and capabilities, as well as establishing organizational-level compliance support systems. Ultimately, this research has provided implications for the responsible use of LLM in professional workflows, balancing efficiency gains from LLM with effective risk management.

%% file: tables/ComplianceExamples.tex
\begin{table}[!ht]
\centering
\caption{Occupation-Specific Examples of LLM-driven Applications}
\renewcommand{\arraystretch}{1.5} 
\setlength{\tabcolsep}{4pt} 
\resizebox{\textwidth}{!}{%
\begin{tabular}{p{3cm} |p{4cm} p{4cm} p{4cm} p{4cm}}
\toprule[1.2pt]
\textbf{Occupation} & \textbf{\makecell{Generating standardized \\ textual content}} & \textbf{\makecell{Extracting insights \\ from  unstructured data}} & \textbf{\makecell{Initial construction \\ of  controlled workflows}} & \textbf{\makecell{Fully automated \\ workflows}} \\ 
\midrule[1.2pt]

Academic & 
Crafting Introductions for Papers, Writing Grant Proposals, and Composing Academic Articles & 
Conducting a Literature Review, Unveiling Contributions and Motivations & 
Standardizing experimental data for scientific research and discovery & 
Automatically generate academic content \\ \cmidrule{1-5}

Public Servants & 
Drafting Policy Documents, Issuing Announcements & 
Summary of Public Opinions, Policy Impact Assessment & 
Policy project planning, Issue Resolution & 
Automatic Response to the Public \\ \cmidrule{1-5}

Health & 
Drafting Medical Reports, Summaries of Patient Records & 
Exploring medical literature for treatment options & 
Managing patient data, Interpreting test results & 
Provide AI diagnosis suggestions \\ \cmidrule{1-5}

Technology & 
Technical documentation drafting, code writing, and annotation. & 
Log analysis, user behavior data analysis & 
Automated testing and model building. & 
Fully Automated Software Development\\ \cmidrule{1-5}

Creative Industry & 
Crafting Live Stream Scripts, Social Media Copy & 
User review sentiment analysis, content creative inspiration, response evaluation & 
Craft a script or sketch outline. & 
Tailored Content Creation \\ \cmidrule{1-5}

Media & 
Press release writing, interview outlines, multilingual translation & 
Automatically generate interview summaries and news summaries & 
Public opinion comment analysis, news planning & 
Compose a News Report \\ \cmidrule{1-5}

Law & 
Drafting of legal documents, contract summaries & 
Search for statutory provisions, analyze case law, and identify patterns of legal risk) & 
Examination of contract clauses, analysis of the facts of the case, and logical hints of arguments & 
Automatically generate defense strategies and precedents, and provide legal advice \\ \cmidrule{1-5}

Commerce & 
Generate financial audit reports, compliance documents, contract documents, and write emails & 
Detection of anomalies in financial statements, automated verification and audit of financial data, and data analysis. & 
Analyze business data and forecast market trends & 
Formulate strategies reports \\ 

\bottomrule[1.2pt]
\end{tabular}}
\setlength{\belowcaptionskip}{3pt}

\end{table}

%% file: tables/ComplianceConcerns.tex
\small
\begin{longtable}{p{0.1\textwidth}|p{0.2\textwidth}|p{0.2\textwidth}|p{0.5\textwidth}}
\caption{Occupation-Specific Compliance Concerns When Using LLM-driven Applications} 
\\ 
\toprule[1.2pt]
\textbf{Occupation} & \textbf{Regulatory Environment} & \textbf{Types of Sensitive Data} & \textbf{Compliance Concerns} \\ 
\midrule[1.2pt]
\endfirsthead

\toprule[1.2pt]
\textbf{Occupation} & \textbf{Regulatory Environment} & \textbf{Types of Sensitive Data} & \textbf{Compliance Concerns} \\ 
\midrule[1.2pt]
\endhead

\multicolumn{4}{r}{\textit{Continued on next page...}} \\ 
\endfoot

\endlastfoot

Academic & 
Subject to the constraints of research ethics, data privacy regulations, intellectual property laws, and institutional policies & 
Depending on the nature of the research, personal data, experimental data, research findings, and other relevant information may be involved & 
\begin{itemize}
    \item Research Data Privacy and Security: Particular emphasis must be placed on ensuring the privacy and security of data, especially when the research involves human subject data.
    \item Academic Integrity and Plagiarism: Concerns arise regarding unintentional non-compliance with academic standards, potentially leading to inadvertent plagiarism and intellectual theft when utilizing LLMs for research and writing assistance.
    \item Intellectual Property Rights of Research Outcomes: Ambiguity surrounding the role of LLMs in the research process may lead to unclear ownership of intellectual property. Knowledge workers, as noted on page 15, express concerns that content generated by LLMs might include data protected by copyright.
\end{itemize}
\\ 
\cmidrule{1-4} 

Public Servants & 
Subject to the constraints of governmental regulations, laws, and professional ethics & 
Depending on the specific position, it may involve personal information of citizens, internal governmental information, etc & 
\begin{itemize}
    \item Information Security and Confidentiality: Concerns over potential leaks when using LLMs to process official information.
    \item Compliance with Policies and Regulations: Ensuring that the use of LLMs adheres to relevant policies and legal frameworks established by government agencies and other public institutions.
    \item Impartiality and Transparency: Avoiding the use of LLMs in a manner that could lead to injustice in public services or a lack of transparency in their implementation.
\end{itemize}
\\ \cmidrule{1-4} 

Health & 
Regulated by medical laws and regulations (e.g., HIPAA), medical ethics, and patient privacy regulations & 
Involves personal privacy information such as the patient's physiology, health status, and medical history, etc & 
\begin{itemize}
    \item Patient Privacy and Data Security: Concerns arise regarding the potential leakage of patients' personal health information when using LLMs to analyze medical records, conduct preliminary diagnoses, or generate patient reports.
    \item Diagnostic and Therapeutic Accuracy and Liability: Concerns that medical advice or diagnostic results provided by LLMs may be inaccurate, and that medical malpractice may result if physicians rely too heavily on LLMs and neglect their professional judgment.
    \item Compliance with Healthcare Industry Norms: Ensure that the use of LLMs is in accordance with standard operating procedures and ethical norms for the medical industry.
\end{itemize}
 \\ \cmidrule{1-4} 

Technology & 
Involves data privacy regulations (such as those applicable when handling user data), open-source license agreements, and position-specific confidentiality provisions & 
Depending on the development project, user data, model training data, and other relevant information may be involved &
\begin{itemize}
    \item Depending on the project under development, user data, model training data, and other related information may be involved. 
    \item Security of the code: Concerns about the potential introduction of security vulnerabilities using LLM-assisted code generation. P8 mentions that the company currently requires that the code cannot be placed in GPT.
    \item Data privacy and security, especially during model training and deployment: Ensure compliance with data privacy regulations when using LLMs to process user data or train models.
    \item Intellectual property and open source licenses: When using LLMs to generate or modify code, you need to comply with the relevant intellectual property rights and open source license agreements. P7 also said that there is no clear path to follow when it comes to data security when it comes to iterative code work with AI.
\end{itemize}
\\ \cmidrule{1-4} 

Creative Industry\& Media & 
Subject to the constraints of copyright law, journalistic ethics, and other regulations & 
It involves the originality of the work and the authenticity of the news & 
\begin{itemize}
    \item Plagiarism and Infringement of Intellectual Property Rights: Concerns that content generated by LLMs may be similar to existing works, leading to accusations of plagiarism. P10 mentioned concerns about using an AI tool for writing novels, which might result in accusations of plagiarism due to the use of fragments from others' works.
    \item False Information and Defamation (especially for journalists): Concerns that news reports generated by LLMs may contain inaccurate information, violating the principle of truthfulness in journalism, and potentially constituting defamation.
    \item Originality and Uniqueness of Works: Concerns that over-reliance on LLMs may affect the originality and personal style of one's works.
\end{itemize}
\\ \cmidrule{1-4} 

Law & 
Strictly governed by laws and regulations, professional ethics of lawyers, and confidentiality provisions & 
Involving confidential client information, proprietary business secrets, and sensitive details pertaining to the case & 
\begin{itemize}
    \item Information Disclosure and Confidentiality Obligations: Concerns that confidential customer information entered into LLMs may be disclosed to third parties or used to train models. P15 refers to concerns about other users finding out about keywords, experimental data results, or important arguments that are communicated to LLMs.
    \item Intellectual Property and Plagiarism Risks: Concerns that LLM-generated legal documents may infringe on the intellectual property rights of others, or that attorneys may fail to adequately review and plagiarize when using LLMs to assist in generating content.
    \item Professional liability and misleading information: concern that reliance on LLM-generated legal advice or arguments may be erroneous or biased, ultimately damaging to clients' interests and personal reputations, attorney P14 argues that the GPT is not a true substitute for an attorney and that the use of misinformation can be challenged by clients.
\end{itemize}
  \\ \cmidrule{1-4} 

Commerce & 
Subject to the regulations and rules imposed by financial regulatory authorities, securities laws, and other legal requirements, as well as professional standards such as auditing standards and accounting regulations & 
Involving the company's business information, financial data, market-sensitive information, customer investment information, etc & 
\begin{itemize}
    \item Insider Trading and Market Manipulation: Concerns that the potential violation of insider trading regulations when sensitive, non-public information is input into LLMs for analysis, or when such activities are detected by competitors using similar tools.
    \item Data Security and Confidentiality: Safeguarding clients' financial information and the company's trade secrets from unauthorized disclosure.
    \item Accuracy and Reliability of LLMs Outputs: Ensuring that recommendations or analysis reports generated using LLMs are accurate and reliable to prevent misleading investors with erroneous information. 
\end{itemize}

\\ \bottomrule[1.2pt]

\end{longtable}